\documentstyle[amssymb]{article}

\begin{document}

\title{A mass generation mechanism for gauge fields}

\author{A. Sevostyanov \\
Department of Mathematical Sciences \\
University of Aberdeen  \\ Aberdeen AB24 3UE, United Kingdom \\
e-mail: seva@maths.abdn.ac.uk \\ phone: +44-1224-272755, fax:
+44-1224-272607}

\maketitle

\begin{abstract}
We suggest a new mass generation mechanism for gauge fields. The
quantum field theory constructed in this paper is nonabelian,
gauge invariant and asymptotically free.
\end{abstract}



\vskip 1cm It is well known that the naive mass generation
mechanism can not be applied in case of the nonabelian Yang-Mills
theory. We recall that if
$F_{\mu\nu}=\partial_{\mu}A_{\nu}-\partial_{\nu}A_{\mu}$ is the
strength tensor of the abelian $U(1)$ gauge field $A_{\mu}$,
$\mu=0,1,2,3$, and $S$ is the usual Maxwell action,
\begin{equation}\label{ab}
S = -\int\frac{1}{4}F_{\mu\nu}F^{\mu\nu}d^4x,
\end{equation}
then one can easily construct the action $S_{m}^0$ generating the
corresponding massive theory (see, for instance, \cite{IZ}, Sect.
3.2.3),
\begin{equation}\label{abm}
S_{m}^0=\int\left(
-\frac{1}{4}F_{\mu\nu}F^{\mu\nu}+\frac{m^2}{2}A_{\mu}A^{\mu}\right)
d^4x.
\end{equation}
Here and thereafter we use the standard convention about
summations and lowering tensor indexes with the help of the
standard metric $g_{\mu\nu}$ of the Minkowski space, $g_{00}=1$,
$g_{ii}=-1$ for $i=1,2,3$, and $g_{ij}=0$ for $i\neq j$.

In contrast to (\ref{ab}) action (\ref{abm}) is not invariant
under gauge transformations
\begin{equation}\label{gauge}
A_\mu \mapsto A_\mu + \partial_\mu \varphi,
\end{equation}
where $\varphi$ is any scalar function on the Minkowski space.
However, if $m\neq 0$ the equations of motion generated by action
(\ref{abm}),
\begin{equation}\label{eq}
\partial^{\mu}F_{\mu\nu}+m^2A_{\nu}=0,
\end{equation}
imply the Lorentz gauge fixing condition
\begin{equation}\label{lor}
\partial^\mu A_\mu =0.
\end{equation}
Condition (\ref{lor}) is obtained by taking the divergence of
equation (\ref{eq}).

Since action (\ref{abm}) is quadratic in $A_\mu$ the field theory
associated to (\ref{abm}) can be easily quantized, the key point
in the quantization procedure being the gauge fixing condition
(\ref{lor}) which automatically appears in the massive case.

If $G$ is a compact simple Lie group the construction of the
massive quantum vector field theory outlined above can not be
carried over in case of the Yang-Mills field associated to $G$.
Formally the action similar to (\ref{abm}) can be written down.
But the new action containing selfinteraction terms turns out to
be nonrenormalizable (see \cite{IZ}, Sect. 12.5.2). The main
obstruction to renormalizability in the nonabelian case is
nonlinearity of the corresponding gauge action. This leads, in
turn, to the fact that the gauge fixing condition similar to
(\ref{lor}) does not automatically appear in the nonabelian case.

Usually one overcomes this difficulty by using spontaneous
symmetry breaking and the Higgs mechanism \cite{H1,H2}. But as a
byproduct of that mechanism one gets extra unwanted boson fields.

In order to avoid the Higgs mechanism for constructing a massive
nonabelian gauge quantum theory we start by slightly modifying
action (\ref{abm}) in such a way that the new action $S_m$ for the
massive field is invariant under gauge transformations
(\ref{gauge}). We introduce the new action $S_m$ by the following
formula
\begin{equation}\label{abmn}
S_{m}=\int\left(
-\frac{1}{4}F_{\mu\nu}F^{\mu\nu}+\frac{m^2}{2}(PA)_{\mu}(PA)^{\mu}\right)
d^4x.
\end{equation}
Here
\begin{equation}\label{pa}
(PA)_{\mu}=A_\mu-\partial_{\mu}\square^{-1}\partial^\nu A_\nu
\end{equation}
is the orthogonal projection of the field $A_\mu$ onto the kernel
of the divergence operator $\partial \cdot A =\partial^\mu A_\mu$,
with respect to the scalar product $<\cdot, \cdot >$ of vector
fields,
$$
<A,B>=\int A_\mu (x) B^\mu (x) d^4x,
$$
and $\square=\partial^\mu \partial_\mu$ is the d'Alambert
operator. In formula (\ref{pa}) it is assumed that an inverse to
the d'Alambert operator acting on the image of $\square$ is fixed.
This can be achieved, for instance, by choosing boundary
conditions for large $x_0=t$. The corresponding boundary
conditions should be also imposed on gauge transformations
$\varphi$ in formula (\ref{gauge}) to insure that each $\varphi$
belongs to the image of $\square^{-1}$ and the identity
\begin{equation}\label{gb1}
\square^{-1}\square \varphi=\varphi
\end{equation}
holds. The particular choice of the operator $\square^{-1}$ is
actually not important. In this paper we shall assume that all
inverse d'Alambert operators are of Feynman type.

Observe that $(PA)_{\mu}$ is the gauge transform (\ref{gauge}) of
$A_{\mu}$ for $\varphi=-\square^{-1}\partial^\nu A_\nu$.

By definition the quantity $(PA)_{\mu}$ is invariant under gauge
transformations (\ref{gauge}). Indeed, if $A_\mu \mapsto A_\mu +
\partial_\mu \varphi$ then
$$
(PA)_\mu \mapsto
A_\mu+\partial_\mu\varphi-\partial_{\mu}\square^{-1}\partial^\nu(A_\nu+\partial_\nu\varphi)=
(PA)_\mu.
$$
$(PA)_{\mu}$ is actually the gauge invariant part of $A_\mu$.
Moreover, $(PA)_{\mu}$ satisfies the Lorentz condition
(\ref{lor}),
\begin{equation}\label{lorpa}
\partial^\mu(PA)_{\mu}=\partial^\mu A_\mu-
\partial^\mu \partial_{\mu}\square^{-1}\partial^\nu A_\nu = 0.
\end{equation}
Note that $(PA)_{\mu}$ linearly depends on $A_\mu$ since the gauge
action is linear in the abelian case.

Since $(PA)_{\mu}$ is invariant under gauge transformations action
(\ref{abmn}) is also gauge invariant.

Observe that the lagrangian density associated to (\ref{abmn}) is
not local. From the first sight this is a serious obstruction for
using action (\ref{abmn}) in quantum field theory. However, if
$$
\partial^\mu A_\mu =0
$$
one has by definition (\ref{pa})
$$
(PA)_{\mu}=A_\mu-\partial_{\mu}\square^{-1}\partial^\nu
A_\nu=A_\mu.
$$
Therefore after imposing gauge condition (\ref{lor}) action
(\ref{abmn}) coincides with action (\ref{abm}) and, hence,
generates the same equations of motion as (\ref{abm}) modulo gauge
transformations. We conclude that action (\ref{abmn}) can be
quantized similarly to (\ref{abm}), and the corresponding Green
functions for {\em local gauge invariant} observables constructed
with the help of (\ref{abmn}) satisfy the locality condition.

From the above discussion we infer that actions (\ref{abm}) and
(\ref{abmn}) are physically equivalent and generate the same
quantum field theory. In this sense action (\ref{abmn}) is a
trivial generalization of (\ref{abm}). But a remarkable property
of the former one is that it has {\em a gauge invariant
renormalizable} counterpart in the nonabelian case!

Formula (\ref{abmn}) is not yet the most suitable for generalizing
to the nonabelian case since it contains the operator $P$ which is
related to the explicit form of the gauge action (\ref{gauge}). In
order to exclude the operator $P$ from formula (\ref{abmn}) we
observe that since $(PA)_\mu$ is a gauge transform of $A_\mu$, and
the strength tensor is gauge invariant, one also has
\begin{equation}\label{ffp}
F_{\mu\nu}=\partial_{\mu}A_{\nu}-\partial_{\nu}A_{\mu}=
\partial_{\mu}(PA)_{\nu}-\partial_{\nu}(PA)_{\mu}.
\end{equation}
Now using (\ref{lorpa}), (\ref{ffp}) and integrating by parts we
obtain the following formula
$$
-\int\frac{1}{2}(\square^{-1}F_{\mu\nu})F^{\mu\nu} d^4x=\int
(PA)_{\mu}(PA)^{\mu} d^4x.
$$
The last relation allows to rewrite action (\ref{abmn}) in an
alternative form,
\begin{equation}\label{abmn1}
S_{m}=-\frac{1}{4}\int\left(
F_{\mu\nu}F^{\mu\nu}+m^2(\square^{-1}F_{\mu\nu})F^{\mu\nu}\right)
d^4x.
\end{equation}
The linearity of the gauge action in the abelian case is actually
not crucial for formula (\ref{abmn1}) since it does not explicitly
contain the operator $P$.

Now it is easy to write down a nonabelian counterpart of action
(\ref{abmn1}). First we fix the notation as in \cite{IZ}. Let $G$
be a compact simple Lie group, $\mathfrak g$ its Lie algebra with
the commutator denoted by $[\cdot, \cdot ]$. We fix a
nondegenerate invariant under the adjoint action scalar product on
$\mathfrak g$ denoted by $\textrm{tr}$ (for instance, one can take
the trace of the composition of the elements of $\mathfrak g$
acting in the adjoint representation). Let $t^a$, $a=1,\ldots
,\textrm{dim}{\mathfrak g}$ be a linear basis of $\mathfrak g$
normalized in such a way that
$\textrm{tr}(t^at^b)=-\frac{1}{2}\delta^{ab}$.

We denote by $A_\mu$ the $\mathfrak g$-valued gauge field
(connection on the Minkowski space),
$$
A_\mu=A_\mu^at^a.
$$
Let $D_\mu$ be the associated covariant derivative,
$$
D_\mu=\partial_\mu-\textrm{g}A_\mu,
$$
where $\textrm{g}$ is a coupling constant, and $F_{\mu\nu}$ the
strength tensor (curvature) of $A_\mu$,
$$
 F_{\mu\nu}=\partial_\mu
A_\nu - \partial_\nu A_\mu -\textrm{g}[A_\mu,A_\nu].
$$

We shall also need a covariant d'Alambert operator $\square_A$
associated to the gauge field $A_\mu$,
$$
\square_A=D_\mu D^\mu.
$$
The covariant d'Alambert operator can be applied to any tensor
field defined on the Minkowski space and taking values in a
representation space of the Lie algebra $\mathfrak g$, the
$\mathfrak g$-valued gauge field $A_\mu$ acts on the tensor field
according to that representation. Note that the operator
$\square_A$ is scalar, i.e. it does not change types of tensors.

Finally recall that the gauge group of $G$-valued functions $g(x)$
defined on the Minkowski space acts on the gauge field $A_\mu$ by
\begin{equation}\label{gaugen}
A_\mu \mapsto \frac{1}{\textrm{g}}(\partial_\mu g)g^{-1}+gA_\mu
g^{-1}.
\end{equation}
The corresponding transformation laws for the covariant derivative
and the strength tensor are
\begin{eqnarray}
D_\mu \mapsto gD_\mu g^{-1}, \label{trcov}\\
F_{\mu\nu} \mapsto gF_{\mu\nu}g^{-1}. \label{trf}
\end{eqnarray}
Formula (\ref{trcov}) implies that the covariant d'Alambert
operator is transformed under gauge action (\ref{gaugen}) as
follows
\begin{equation}\label{trdal}
\square_A \mapsto g\square_A g^{-1}.
\end{equation}
In the last formula we assume that the gauge group acts on tensor
fields according to the representation of the group $G$ induced by
that of the Lie algebra $\mathfrak g$.

From formulas (\ref{trf}) and (\ref{trdal}) it follows that the
natural nonabelian analogue of action (\ref{abmn1}) invariant
under gauge action (\ref{gaugen}) is
\begin{equation}\label{m}
S_{m}^G=\frac{1}{2}\int\textrm{tr} \left(
F_{\mu\nu}F^{\mu\nu}+m^2(\square^{-1}_AF_{\mu\nu})F^{\mu\nu}\right)
d^4x,
\end{equation}
where in the expression $(\square^{-1}_AF_{\mu\nu})$ we assume
that $F_{\mu\nu}$ is in the adjoint representation of $\mathfrak
g$ and the scalar operator $\square^{-1}_A$ acts on $F_{\mu\nu}$
componentwise. The first term in (\ref{m}) is simply the action of
the Yang-Mills field.

Now a new problem immediately appears. Due to nonlinearity of the
gauge action in the nonabelian case it is impossible to make the
lagrangian density associated to action (\ref{m}) local by
imposing a gauge fixing condition. However, the appearance of
nonlocal quantities in quantum field theory is not a new fact. One
should recall, first of all, the Faddeev-Popov determinant
\cite{FP} that can be made local using additional anticommuting
ghost fields. In this paper we are going to apply a similar trick
to action (\ref{m}).

Similarly to the case of pure Yang-Mills field (see \cite{IZ},
Sect. 12.2; \cite{FS}, Ch. 3, \S 3) let us first formally write
down the expression for the generating functional $Z(J)$ of the
Green functions associated to action (\ref{m}) via a Feynman path
integral,
\begin{eqnarray}\label{act}
Z(J)=\int{\mathcal D}(A){\mathcal D}(\eta){\mathcal
D}(\overline{\eta})\exp \{ i \int [  \textrm{tr} \big(
\frac{1}{2}F_{\mu\nu}F^{\mu\nu}+\frac{m^2}{2}(\square^{-1}_AF_{\mu\nu})F^{\mu\nu}
\\
\qquad \qquad \qquad \qquad \qquad \qquad \qquad +2\overline{\eta}
({\mathcal M} \eta)-2J^\mu A_\mu \big)+{\mathcal F}(A) ] d^4x \} ,
\nonumber
\end{eqnarray}
where $J^\mu$ is the source for $A_\mu$ taking values in the Lie
algebra $\mathfrak g$, ${\mathcal F}(A)$ is a gauge fixing term,
${\mathcal M}$ the corresponding Faddeev-Popov operator,
$\overline{\eta}, \eta$ are the anticommuting scalar ghost fields
taking values in the adjoint representation of $\mathfrak g$
(Faddeev-Popov ghosts). Here and later we always assume that the
measure $\mathcal D$ in Feynman path integrals is properly
normalized.

The key observation is that the nonlocal part of action (\ref{m})
is the quadratic form of the operator $\square^{-1}_A$. Therefore
to make the expression in the exponent in formula (\ref{act})
explicitly local it is natural to use a formula for Gaussian
integrals relating the exponent of the quadratic form of the
operator and the exponent of the quadratic form of the inverse
operator. For operator $K$ acting on scalar fields that formula
takes the following form (\cite{FS}, Ch. 2, \S 6)
\begin{eqnarray}\label{inv}
\exp \{ \frac{i}{2} \int (K^{-1}\varphi)\varphi d^4x\}= \qquad
\qquad \qquad \qquad \qquad \qquad \qquad \qquad
\\
\qquad \qquad  =\textrm{det}(K)^{\frac{1}{2}}
  \int{\mathcal D}(\psi)\exp \{  i\int( -\frac{1}{2}(K\psi)\psi
+\varphi\psi ) d^4x\}. \nonumber
\end{eqnarray}

Since in formula (\ref{act}) the operator $\square^{-1}_A$ acts on
the skew-symmetric (2,0)-type tensor field $F_{\mu\nu}$, in order
to apply a formula similar to (\ref{inv}) to the expression
\begin{equation}\label{nonloc}
\exp \{ i \int
\textrm{tr}
\big(\frac{m^2}{2}(\square^{-1}_AF_{\mu\nu})F^{\mu\nu}\big) d^4x
\}
\end{equation}
one has to introduce a new skew-symmetric (2,0)-type tensor field
$\Phi_{\mu\nu}$ (bosonic ghost) with values in the adjoint
representation of the Lie algebra $\mathfrak g$.

Using the ghost field $\Phi_{\mu\nu}$ and formula (\ref{inv}) we
can rewrite expression (\ref{nonloc}) in an alternative form,
\begin{eqnarray}\label{loc1}
\exp \{ i \int\textrm{tr}
\big(\frac{m^2}{2}(\square^{-1}_AF_{\mu\nu})F^{\mu\nu}\big) d^4x
\}= \qquad \qquad \qquad \qquad \qquad \qquad \qquad \qquad
\\
=(\textrm{det}\square_A|_{\Omega^3({\mathfrak g})})^{\frac{1}{2}}
  \int{\mathcal D}(\Phi)\exp \{  i\int \textrm{tr}
  \big( -\frac{1}{8}(\square_A\Phi_{\mu\nu})\Phi^{\mu\nu}
+\frac{m}{2}\Phi_{\mu\nu}F^{\mu\nu} \big) d^4x\}, \nonumber
\end{eqnarray}
where the operator $\square_A$ is assumed to act componentwise in
the space $\Omega^2({\mathfrak g})$ of skew-symmetric (2,0)-type
tensors with values in the adjoint representation of $\mathfrak
g$. The numerical factors in the r.h.s. of the last formula appear
since the Feynman path integral in (\ref{loc1}) is only taken with
respect to the linearly independent components $\Phi_{\mu\nu}$,
$\mu < \nu$ of the skew-symmetric tensor $\Phi_{\mu\nu}$.

The r.h.s. of (\ref{loc1}) still contains a nonlocal term,
$(\textrm{det}\square_A|_{\Omega^2({\mathfrak
g})})^{\frac{1}{2}}$. But since $\Omega^2({\mathfrak g})$ is the
direct sum of six copies of the space $\Omega^0({\mathfrak g})$ of
scalar fields with values in the adjoint representation of
$\mathfrak g$ and the operator $\square_A$ acts componentwise in
$\Omega^2({\mathfrak g})$ the determinant
$\textrm{det}\square_A|_{\Omega^2({\mathfrak g})}$ is equal to
$(\textrm{det}\square_A|_{\Omega^0({\mathfrak g})})^6$, and
$$
(\textrm{det}\square_A|_{\Omega^2({\mathfrak
g})})^{\frac{1}{2}}=(\textrm{det}\square_A|_{\Omega^0({\mathfrak
g})})^3.
$$
Therefore recalling the Faddeev-Popov trick (see \cite{FS}, Ch. 3,
\S 3; \cite{IZ}, Sect. 12.2.2) and introducing six scalar
anticommuting ghost fields $\eta_i, \overline{\eta}_i$, $i=1,2,3$
taking values in the adjoint representation of $\mathfrak g$ we
can express the square root of the determinant,
$(\textrm{det}\square_A|_{\Omega^2({\mathfrak
g})})^{\frac{1}{2}}$, in the following form
\begin{eqnarray}\label{detloc}
(\textrm{det}\square_A|_{\Omega^2({\mathfrak
g})})^{\frac{1}{2}}=(\textrm{det}\square_A|_{\Omega^0({\mathfrak
g})})^3=\qquad \qquad \qquad \qquad \qquad \qquad
\\
=\int\prod_{i=1}^3{\mathcal D}(\eta_i){\mathcal
D}(\overline{\eta}_i)\exp \{ i \int  2\textrm{tr}
\big(\sum_{i=1}^3 \overline{\eta}_i (\square_A \eta_i) \big) d^4x
\}. \nonumber
\end{eqnarray}

Now substituting (\ref{loc1}) and (\ref{detloc}) into into
(\ref{act}) we derive an expression for the generating function
$Z(J)$ that only contains local interaction terms and suitable for
developing perturbation theory,
\begin{eqnarray}
Z(J)=\int{\mathcal D}(A){\mathcal D}(\Phi){\mathcal
D}(\eta){\mathcal D}(\overline{\eta})\prod_{i=1}^3{\mathcal
D}(\eta_i){\mathcal D}(\overline{\eta}_i) \times \qquad \qquad
\qquad \nonumber
\\*
\times \exp \left\{ i \int [ \textrm{tr} \left(
\frac{1}{2}F_{\mu\nu}F^{\mu\nu}-\frac{1}{8}(\square_A\Phi_{\mu\nu})\Phi^{\mu\nu}
+\frac{m}{2}\Phi_{\mu\nu}F^{\mu\nu}+ \right. \right.
\label{actloc} \\*
\left. \left. +2\overline{\eta} ({\mathcal M}
\eta) +2\sum_{i=1}^3 \overline{\eta}_i (\square_A \eta_i)-2J^\mu
A_\mu \right)+{\mathcal F}(A) ] d^4x \right\} . \nonumber
\end{eqnarray}

Formula (\ref{actloc}) looks a little bit complicated. But it is
equivalent to the compact expression (\ref{act}). Note that when
deriving (\ref{actloc}) starting from (\ref{act}) we only used
tricks with Gaussian Feynman path integrals. From (\ref{act}) it
also follows that when $m=0$ the generating function $Z(J)$ is
reduced to that of the quantized Yang-Mills theory.

Now we can discuss renormalization of the lagrangian in formula
(\ref{actloc}). In order to do that we have to first fix a gauge,
i.e. we have to choose a gauge fixing term ${\mathcal F}(A)$ and
specify the Faddeev-Popov operator ${\mathcal M}$.

We shall see in a moment that for our purposes the most convenient
choice of ${\mathcal F}(A)$ is
\begin{equation}\label{F}
{\mathcal F}(A)=\textrm{tr}(\partial^\mu A_\mu \partial^\nu A_\nu
+ m^2\partial^\mu A_\mu (\square^{-1}
\partial^\nu A_\nu)).
\end{equation}
Since the r.h.s. of (\ref{F}) only depends on the divergence
$\partial^\mu A_\mu$ of $A_\mu$ the corresponding Faddeev-Popov
operator has the same form as in case of the Lorentz gauge (see
\cite{FS}, Ch. 3, \S 3),
\begin{equation}\label{fp}
{\mathcal M}=\partial^\mu D_\mu.
\end{equation}

Note that gauge fixing condition (\ref{F}) contains a nonlocal
term. But since the nonlocal term only depends on the longitudinal
component of the gauge field $A_\mu$ the nonlocal part in
(\ref{F}) does not generate extra nonlocal terms in the
expressions for gauge invariant physical observables (see
\cite{FS}, Ch. 3, \S 3 for discussion of a similar situation in
case of pure Yang-Mills field).

Finally combining (\ref{actloc}), (\ref{F}) and (\ref{fp}) we
arrive at the following formula for $Z(J)$
\begin{eqnarray}
Z(J)=\int{\mathcal D}(A){\mathcal D}(\Phi){\mathcal
D}(\eta){\mathcal D}(\overline{\eta})\prod_{i=1}^3{\mathcal
D}(\eta_i){\mathcal D}(\overline{\eta}_i) \times \qquad \qquad
\qquad \qquad \qquad \nonumber
\\
\times \exp \left\{ i \int \textrm{tr} \left(
\frac{1}{2}F_{\mu\nu}F^{\mu\nu}-\frac{1}{8}(\square_A\Phi_{\mu\nu})\Phi^{\mu\nu}
+\frac{m}{2}\Phi_{\mu\nu}F^{\mu\nu}-2J^\mu
A_\mu + \right. \right. \qquad \label{actloc1}\\
\left. \left. +2\overline{\eta} (
\partial^\mu D_\mu
\eta)+2\sum_{i=1}^3 \overline{\eta}_i (\square_A
\eta_i)+\partial^\mu A_\mu \partial^\nu A_\nu + m^2\partial^\mu
A_\mu (\square^{-1}
\partial^\nu A_\nu) \right) d^4x \right\} . \nonumber
\end{eqnarray}

By simple dimensional counting the quantum field theory with
generating function (\ref{actloc1}) is renormalizable. Moreover,
using dimensional regularization and the BRST technique one can
show, similarly to the case of the Yang-Mills field (see
\cite{IZ}), that gauge invariance is preserved by the
renormalization. The details of those proofs and of other
calculations that we are going to discuss can be found in
\cite{S}.

The particular choice (\ref{F}) of the gauge fixing term is
convenient since in this case the free propagator
$D_{\mu\nu}^{ab}$ of the gauge field $A_\mu$ has a very simple
form,
$$
D_{\mu\nu}^{ab}=i\delta^{ab}\frac{g_{\mu\nu}}{\square+m^2}.
$$
Therefore the gauge field $A_\mu$ becomes massive. However, the
theory may also contain some massless particles corresponding to
the ghost fields.

The coupling constant renormalization calculated at the one--loop
order using dimensional regularization coincides with that for the
pure Yang-Mills field (see \cite{IZ}),
\begin{equation}\label{gren}
\textrm{g}_0=\textrm{g}\mu^{\frac{\varepsilon}{2}}\left(
1-\frac{\textrm{g}^2}{16\pi^2}\frac{11}{6}C\frac{2}{\varepsilon}\right),
\end{equation}
where $\textrm{g}_0$ is the bare coupling constant, $\varepsilon
\rightarrow 0$ when the regularization is removed, $\mu$ is an
arbitrary mass scale, and the constant $C$ is defined by
$$
C\delta_{ab}=\sum_{c,d}C^{cd}_aC^{cd}_b
$$
with the help of the structure constants $C^{ab}_c$ of the Lie
algebra $\mathfrak g$, $[t^a,t^b]=C^{ab}_ct^c$.

It is easy to see that the coupling constant renormalization in
the theory with generating function (\ref{actloc1}) calculated
within the dimentional regularization framework coincides with the
coupling constant renormalization for the pure Yang-Mills theory
to all orders of perturbation theory. Indeed, by the results of
\cite{th} in the case of dimensional regularization the coupling
constant renormalization is independent of the mass for
dimensional reasons. Therefore the coupling constant
renormalization in the theory with generating function
(\ref{actloc1}) is the same as in the massless case, i.e. in case
of the Yang-Mills field.

One can also calculate the mass renormalization at the one--loop
order,
\begin{equation}\label{mren}
m_0=m\mu^{\frac{\varepsilon}{2}}\left(
1-\frac{\textrm{g}^2}{16\pi^2}\frac{11}{6}C\frac{2}{\varepsilon}\right),
\end{equation}
where $m_0$ is the bare mass.

The wave function renormalization at the one--loop order has the
same form as in case of the Yang-Mills field,
\begin{equation}\label{Aren}
Z_A=1+\frac{\textrm{g}^2}{16\pi^2}\frac{5}{3}C\frac{2}{\varepsilon}.
\end{equation}

From (\ref{gren}), (\ref{mren}) and (\ref{Aren}) one can get the
corresponding renormalization group coefficients at the one-loop
order (see \cite{R}, Sect. 4.5),
\begin{eqnarray}
\beta(\textrm{g})=\mu\frac{\partial
\textrm{g}}{\partial\mu}=-\frac{\textrm{g}^3}{16\pi^2}\frac{11}{3}C, \qquad \quad \\
\gamma_m(\textrm{g})=\mu\frac{\partial \ln
m}{\partial\mu}=-\frac{\textrm{g}^2}{16\pi^2}\frac{11}{6}C, \qquad  \label{gm} \\
\gamma_d(\textrm{g})=\frac{1}{2}\mu\frac{\partial \ln
Z_A}{\partial\mu}=-\frac{\textrm{g}^2}{16\pi^2}\frac{5}{3}C. \quad
\end{eqnarray}

The Fourier transform $\Gamma(p_1,\ldots, p_n ,m,\textrm{g},\mu)$
of any one-particle irreducible Green function with engineering
dimension $d$ is transformed under rescaling of the momenta as
follows (see \cite{R}, Sect. 4.6)
\begin{eqnarray}\label{gamma}
\Gamma(sp_1,\ldots, sp_n ,m,\textrm{g},\mu)= \qquad \qquad \qquad \qquad \qquad \qquad \qquad \qquad \qquad  \\
\qquad \qquad  =s^{d}\Gamma(p_1,\ldots, p_n
,m(s),\textrm{g}(s),\mu)\exp\left\{
-n\int_1^s\frac{ds'}{s'}\gamma_d(\textrm{g}(s'))\right\},
\nonumber
\end{eqnarray}
where $s>0$ is a scale for the momenta. In equation (\ref{gamma})
$\textrm{g}(s)$ and $m(s)$ are the solutions to the following
equations
\begin{eqnarray}
s\frac{\partial \textrm{g}(s)}{\partial s}=\beta(\textrm{g}(s)),\quad \textrm{g}(1)=\textrm{g}; \qquad \qquad \qquad  \label{ug} \\
s\frac{\partial m(s)}{\partial
s}=m(s)(\gamma_m(\textrm{g}(s))-1),\quad m(1)=m. \label{um}
\end{eqnarray}

Applying standard renormalization group arguments (see, for
instance, \cite{IZ}, \cite{R}) we have $\textrm{g}(s)\rightarrow
0$ when $s\rightarrow \infty$ since the beta function
$\beta(\textrm{g})$ is negative in (\ref{ug}), and the theory with
generating function (\ref{actloc1}) is asymptotically free.

Similarly, $m(s)\rightarrow 0$ when $s\rightarrow \infty$ since in
that case $\textrm{g}(s)\rightarrow 0$ as we have just observed,
and hence by (\ref{gm}) $\gamma_m(\textrm{g}(s))-1$ is negative in
(\ref{um}) for large $s$. Therefore the mass can be neglected for
large momenta.

In conclusion we remark that the mass term in action (\ref{m}) can
be regarded as the result of a renormalization of the Yang-Mills
action. Such possibility was discussed in \cite{F}. However, the
mass term can not be observed perturbatively at large momenta. We
also note that there should be no serious problems with
generalizing to our theory the program of rigorous construction of
Green functions for the quantized Yang-Mills field outlined in
\cite{MRS}.

One can also add to action (\ref{m}) a fermionic part coupled to
the gauge field $A_\mu$ in the minimal way.

\vskip 0.5cm

{\large \bf Acknowledgements.} When this work was completed the
author has received information that the gauge invariant mass term
that appears in formula (\ref{abmn1}) in the abelian case was
considered in \cite{J}, and action
 (\ref{m}) in the nonabelian case was recently introduced and studied in
\cite{C}. Paper \cite{C} also contains an extensive list of
references on dynamical mass generation in the Yang-Mills theory.

Another action describing massive gauge fields was proposed in
\cite{L}.

Prof. R. Jackiw also communicated to me that in case of three
dimensions the mass generation mechanism similar to the one
introduced in this paper was suggested in \cite{Jw}. The author
would like to thank him, David Dudal, Amitabha Lahiri and Jean-Luc
Jacquot for pointing out the references quoted above.


\begin{thebibliography}{99}
\bibitem{C} Capri, M. A. L., Dudal, D., Gracey, J. A., Lemes, V. E. R.,
Sobreiro, R. F., Sorella, S. P., Verschelde, H., Study of the
gauge invariant, nonlocal mass operator $\textrm{Tr}\int
d^4xF_{\mu\nu}(D^2)^{-1}F_{\mu\nu}$ in Yang-Mills theories, {\em
Phys. Rev.} {\bf D 72} (2005), 105016.

\bibitem{F} Faddeev, L. D., Mass in quantum Yang-Mills theory
(comment on a Clay millenium problem), {\em Bull. Braz. Math.
Soc., New Series} {\bf 33(2)} (2002), 201--212.

\bibitem{FP} Faddeev, L. D., Popov V. N., Feynman
diagrams for the Yang-Mills field, {\em Phys. Lett.}  {\bf B 25}
(1967), 29.

\bibitem{FS} Faddeev, L. D., Slavnov, A. A., Gauge fields.
Introduction to quantum theory, {\em Frontiers in Physics, 83}
Addison-Wesley (1991).

\bibitem{H1} Higgs, P. W., Broken symmetries and the masses of
gauge bosons, {\em Phys. Rev. Lett.} {\bf 13} (1964), 508--509.

\bibitem{H2} Higgs, P. W., Spontaneous symmetry breakdown without
massless bosons, spontaneous symmetry breakdown without massless
bosons, {\em Phys. Rev. (2)} {\bf 145} (1966), 1156--1163.

\bibitem{th} 't Hooft, G., Dimensional regularization and the
renormalization group, {\em Nucl. Phys.} {\bf B 61} (1973),
455--468.

\bibitem{IZ} Itzykson, C., Zuber, J.-B., Quantum Field Theory,
McGraw-Hill (1980).

\bibitem{Jw} Jackiw, R., Pi, S.-Y., Seeking an even-parity mass
term for 3-D gauge theory, {\em Phys. Lett.} {\bf B 403} (1997),
297.

\bibitem{J} Jacquot, J. L., Path integral regularization of QED by
means of Stueckelberg fields, {\em Phys. Lett.} {\bf B 631} (2005)
83.

\bibitem{L} Lahiri, A., Generating Vector Boson Masses, preprint
hep-th/9301060; The Dynamical Nonabelian Two-Form: BRST
Quantization, {\em Phys. Rev.} {\bf D55} (1997), 5045-5050;
Renormalizability of the Dynamical Two-Form, {\em Phys. Rev.} {\bf
D63} (2001), 105002.

\bibitem{MRS} Magnen, J., Rivasseau, V., S\'{e}n\'{e}or, R.,
Construction of $YM_4$ with an infrared cutoff, {\em Comm. Math.
Phys.} {\bf 155} (1993), 325--383.

\bibitem{R} Ramond, P., Field theory: a modern primer,
Addison-Wesley (1989).

\bibitem{S} Sevostyanov, A., Nonlocal lagrangians and mass
generation for gauge fields, preprint hep-th/0605051.


\end{thebibliography}
\end{document}